\documentclass{article}

\usepackage{arxiv}

\usepackage[utf8]{inputenc} 
\usepackage[T1]{fontenc}    
\usepackage{hyperref}       
\usepackage{url}            
\usepackage{booktabs}       
\usepackage{amsfonts}       
\usepackage{nicefrac}       
\usepackage{microtype}      
\usepackage{lipsum}
\usepackage{amsmath}
\usepackage{graphicx}
\usepackage{amssymb}
\usepackage{float}

\newcommand{\beginsupplement}{%
        \setcounter{table}{0}
        \renewcommand{\thetable}{S\arabic{table}}%
        \setcounter{figure}{0}
        \renewcommand{\thefigure}{S\arabic{figure}}%
        \setcounter{section}{0}
        \renewcommand{\thesection}{S\arabic{section}}%
        \setcounter{equation}{0}
        \renewcommand{\theequation}{S\arabic{equation}}%
     }

\begin{document}

\begin{flushleft}
{\Large
\textbf\newline{\textbf{Zernike mode rescaling extends capabilities of adaptive optics for microscopy}} 
}
\newline
\\
Jakub Czuchnowski\textsuperscript{1,*,\S},
Robert Prevedel\textsuperscript{1,**}
\\
\bigskip
\textsuperscript{1} Cell Biology and Biophysics Unit, European Molecular Biology Laboratory, Heidelberg, Germany
\\
\textsuperscript{\S} Collaboration for joint PhD degree between EMBL and Heidelberg University, Faculty of Biosciences, Germany
\\
\bigskip
* jakub.czuchnowski@embl.de\\
** robert.prevedel@embl.de

\end{flushleft}

\begin{abstract}
Zernike polynomials are widely used mathematical models of experimentally observed optical aberrations. Their useful mathematical properties, in particular their orthogonality, make them a ubiquitous basis set for solving various problems in beam optics. Thus they have found widespread use in adaptive optics realizations that are used to correct wavefront aberrations. However, Zernike aberrations lose their orthogonality when used in combination with Gaussian beams, which are omnipresent in real-world optical applications. As a consequence, Zernike aberrations in Gaussian beams start to cross-couple between each other, a phenomenon that does not occur for Zernike aberrations in plane waves. Here, we describe how the aberration radius influences this cross-coupling of Zernike aberrations. Furthermore, we propose that this effect can actually be harnessed to allow efficient compensation of higher-order aberrations using only low-order Zernike modes. This finding has important practical implications, as it suggests the possibility of using adaptive optics devices with low element numbers to compensate aberrations which would normally require more complex and expensive devices.

\end{abstract}

\section{Introduction}
Adaptive optics (AO) allowed breakthrough discoveries in astronomy by enabling telescopic observation through difficult atmospheric conditions \cite{hardy1998adaptive}. These breakthroughs were enabled by important developments in understanding and tackling optical aberrations introduced into light during its propagation. In recent years, developments of AO for microscopic imaging allowed for high resolution visualisation of structures deep in scattering materials (e.g. biological tissues)\cite{Booth:14,Ji:17}. These, however, require dedicated hardware different from the one used in astronomy, but maybe even more importantly, also different theoretical frameworks suited to the conditions used in microscopy (e.g. use of Gaussian beams for laser microscopy). Currently, the dominant theoretical model of optical aberrations in microscopy are Zernike polynomials due to their orthogonal nature and isomorfisms to experimentally observable aberrations \cite{booth2007adaptive}.

However, previous work has shown that Zernike aberrations are not orthogonal in the case of Gaussian beams and can display significant cross-coupling as described both by the Strehl ratio approach \cite{mahajan1994zernike,mahajan1995zernike} as well as by evaluating coupling into higher-order Laguerre-Gauss (LG) modes \cite{Czuchnowski:21}. While more optimised basis sets for experimental AO have indeed been proposed \cite{debarre2009image}, here we pursue an alternative approach that actually harnesses the non-ortogonality of Zernike aberrations.

In particular, we show that by manipulating the Zernike mode size with relation to the Gaussian beam diameter (e.g. by changing the ratio between the active aperture of an AO element and the size of the beam) it is possible to strongly enhance cross-coupling properties of Zernike aberrations which could be used in practice to increase the correction capabilities of commonly used AO elements.

\begin{figure*}[h]
\includegraphics[width=16cm]{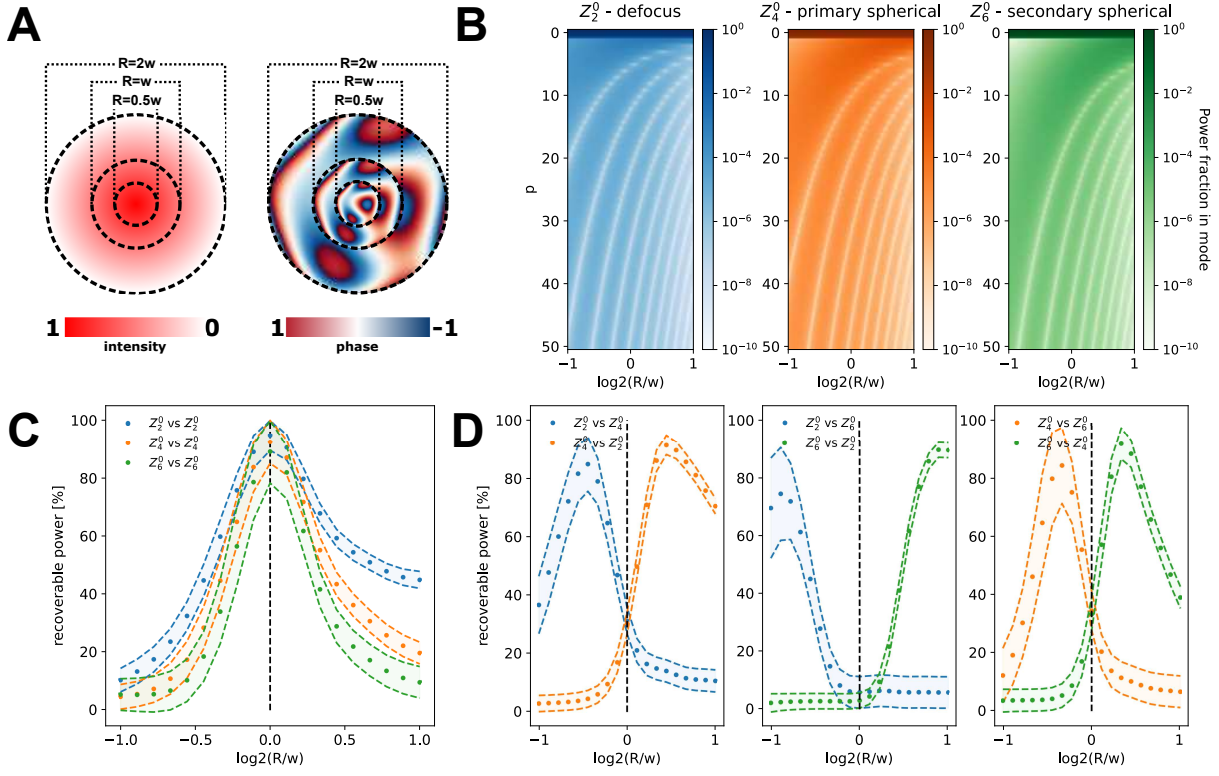}
\centering
\caption{\textbf{A} Cartoon representation of the wavefront dependence on the aberration radius (R). Left: Intensity cross-section. Right: Wavefront amplitude. \textbf{B} Dependence of the power coupling into higher order LG-modes on the ratio between the aberration and beam radii (R/w). \textbf{C} Dependence of the self-compensation in LG-space on the R/w ratio using the first 25 LG modes. The shaded area represents the uncertainty range (which originates from the finite subset of LG modes considered) between the lower and upper bounds calculated using \textbf{Equation 8} and \textbf{Equation S10} (see \textbf{Supplement 1 Section S3} for details). The dots represent the expected value. \textbf{D} Dependence of cross-compensation between different aberrations on the R/w ratio using the first 25 LG modes.} 
\label{fig:1}
\end{figure*}

\section{Effects of aberration radius on coupling between Zernike aberrations and Laguerre-Gauss modes}

Laguerre-Gauss beams are inherently orthogonal and therefore do not cross-couple between each other in free space propagation. However, Zernike type aberrations are capable of inducing energy coupling between different LG-beams \cite{Bond:11}. The coupling coefficient between LG-modes induced by a particular $Z^m_n$ aberration can be decribed as:
\small
\begin{equation}
    k^{n,m}_{p,l,p',l'}=\int_A LG_{p,l}\exp(ikZ^m_n)LG_{p',l'}^*dA
    \label{eq:knmplpl}
\end{equation}
\normalsize

where:
\small
\begin{equation}
Z^m_n(r,\phi) = \begin{cases}
    A^m_n R^m_n(r)\cos(m\phi)  & \text{for $m\geqslant0$}\\
    A^m_n R^{|m|}_n(r)\sin(|m|\phi) & \text{for $m<0$},\\
    \end{cases}   
\end{equation}
\normalsize
\small
\begin{equation}
{LG_{lp}}(r,\phi ,z,k)=C_{lp}^{LG}\left({\frac {r{\sqrt {2}}}{w(z)}}\right)^{\!|l|}L_{p}^{|l|}\!\left({\frac {2r^{2}}{w^{2}(z)}}\right)\exp(-il\phi )G(r,z,k),
\end{equation}
\normalsize

where k denotes the wavenumber and A is the pupil area over which the beam is integrated (see \textbf{Supplement 1 Section S1} for full definitions). In the weak aberration regime the phase aberration can be expressed as:
\small
\begin{equation}
    \exp(ikZ^m_n)\approx1+ikZ^m_n.
\end{equation}
\normalsize
This enables \textbf{Equation \ref{eq:knmplpl}} to be solved analytically \cite{Bond:11}:
\small
\begin{equation}
    k^{n,m}_{p,l,p',l'}=\int_0^{2\pi} \int_0^R LG_{p,l}LG_{p',l'}^*(ikZ^m_n)rdrd\phi=\delta_{p,p'} \delta_{l,l'}+ I_{\phi}I_r
    \label{eq:knmplpl2}
\end{equation}
\normalsize

where $I_\phi$ and $I_r$ are the azimuthal and radial integral respectively. The azimuthal part ($I_\phi$) determines the coupling condition (see \cite{Bond:11,Bond:14,Czuchnowski:21} and \textbf{Supplement 1 Section S2} for details) and provides a mapping between particular groups of Zernike aberrations and LG-modes. However, the radial integral ($I_r$) determines coupling within the groups of LG-modes determined by \textbf{Equation S9} effectively regulating the degree of cross-compensation \cite{Bond:11,Bond:14}:

\small
\begin{multline}
\label{eq:Ir}
    [I_r]^{n,m}_{p,l,p',l'}=A^m_n\frac{ik}{\pi}\sqrt{p!p'!(p+|l|)!(p'+|l'|)!}\exp(i\Delta o \psi) \\
    \times \sum_{i=0}^p \sum_{j=0}^{p'} \sum_{h=0}^{\frac{1}{2}(n-m)} \frac{(-1)^{i+j+h}}{(p-i)!(p'-j)!(|l|+i)!(|l'|+j)!i!j!}\frac{1}{\hat{R}^{\frac{1}{2}(n-2h)}} \times \\
    \frac{(n-h)!}{(\frac{n+m}{2}-h)!(\frac{n-m}{2}-h)!h!}\gamma(i+j-h+\frac{1}{2}(|l|+|l'|+n)+1,\hat{R})
\end{multline}
\normalsize

where $A_n^m$ is the amplitude of the Zernike aberration, $\psi$ is the Gouy phase described by \textbf{Equation S7}, $\hat{R}=\frac{2R^2}{w^2}$ encodes the ratio between the Zernike radius ($R$) and the beam diameter ($w$), $\Delta o=2p+l-2p'-l'$ is the difference in orders between incident and coupled LG-modes and $\gamma(a,x)=\int_0^x t^{a-1}e^{-t}dt$ is the lower incomplete gamma function. As can be appreciated from \textbf{Equation \ref{eq:Ir}} the relation between the coupling coefficient and the ratio between the Zernike radius and the beam diameter ($\hat{R}$) is a complex one and does not allow for straightforward analytical analysis. Therefore we have used this framework to numerically evaluate the effects of aberration radius on the beam properties. One can appreciate that the coupling distribution into higher order LG-modes strongly depends on the ratio between the aberration and beam radii (R/w, \textbf{Figure \ref{fig:1}B, Supplementary Figure S1A}). To check whether this has an effect on Zernike cross-compensation \cite{Czuchnowski:21} we calculate the recoverable power fraction ($P_{rec}$) for beams aberrated with different R/w ratios.

\small
\begin{equation}
   P_{rec}=\frac{\sum_i \big(k_i^{n_p,m_p}(\frac{R_p}{w})+\alpha k_i^{n_a,m_a}(\frac{R_a}{w})\big)\overline{\big(k_i^{n_p,m_p}(\frac{R_p}{w})+\alpha k_i^{n_a,m_a}(\frac{R_a}{w})\big)}}{1-\big|k_{0,0,0,0}^{n_p,m_p}(\frac{R_p}{w})\big|^2}
\end{equation}
\normalsize

where $\alpha$ denotes the optimal correction amplitude which can be calculated from \textbf{Equation S11}, $i=(p,l,p',l')$ excluding $(p',l')=(0,0)$. By setting $(n_a,m_a)=(n_p,m_p)$ we explore the self-compensation of Zerkine aberration with different R/w ratios. We show that the recoverable power decreases (\textbf{Figure \ref{fig:1}C, Supplementary Figure S1B}) which implies that rescaled Zernike modes (with $R/w \neq 1$) are able to compensate the native Zernike aberrations (with $R/w=1$) only to a limited degree (which is intuitive since rescaling reduces the mode self-similarity which can be observed in \textbf{Figure \ref{fig:1}A}). On the other hand, the recoverable power between different aberrations ($(n_a,m_a)\neq(n_p,m_p)$) increases which facilitates stronger cross compensation between the modes (\textbf{Figure \ref{fig:1}D, Supplementary Figure S1C}). Interestingly, reducing the R/w ratio allows Zernike modes to more efficiently cross-compensate higher-order aberrations which has important practical implications as it suggests it's possible to extend aberration correction capabilities of AO elements (e.g. deformable mirrors) beyond their specified limitations.

\begin{figure*}[h]
\includegraphics[width=16cm]{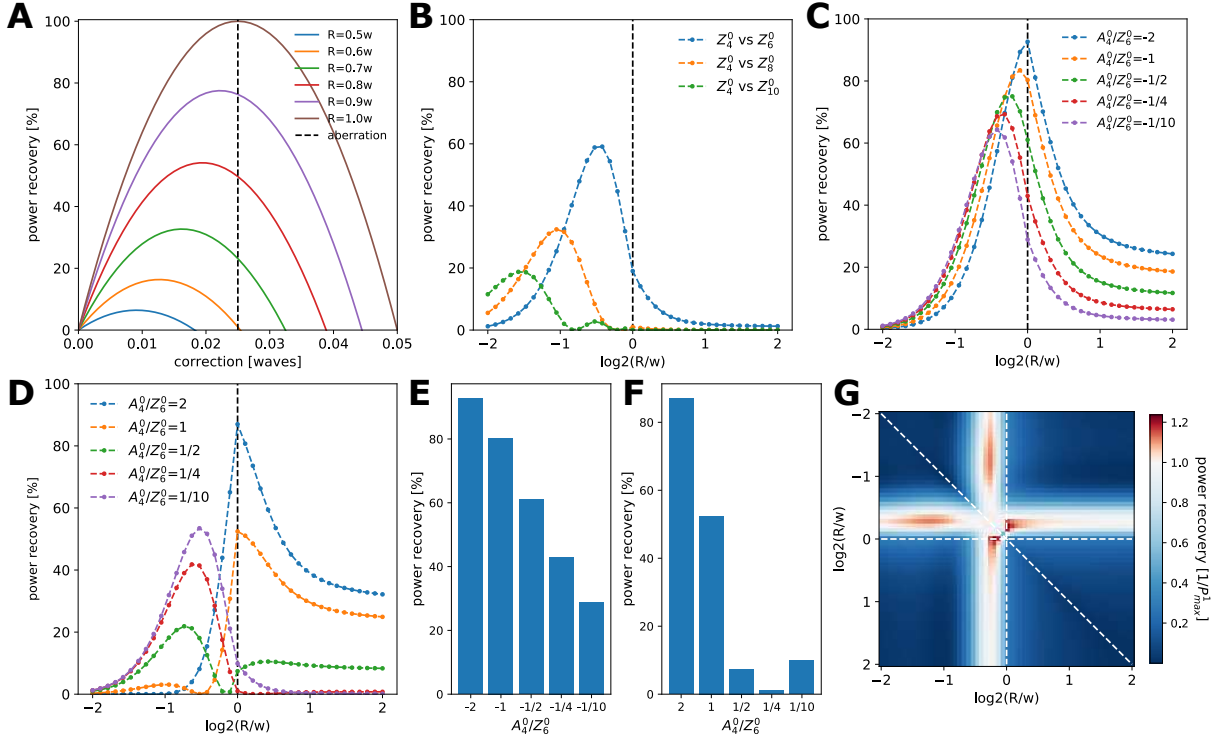}
\centering
\caption{\textbf{A} Self-compensation of $Z^0_4$ (spherical aberration). \textbf{B} Power recovery for 3 higher order Z-aberrations ($Z^0_6$, $Z^0_8$, $Z^0_{10}$) when compensated with the lower order $Z^0_4$ mode (shperical aberration). \textbf{C,D} Power recovery for a combined $Z^0_4$+$Z^0_6$ aberration with different negative (\textbf{C}) and positive (\textbf{D})  mixing ratios ($A^0_4/A^0_6$) when compensated with a $Z^0_4$ mode. \textbf{E,F} Power recovery for a combined $Z^0_4$+$Z^0_6$ aberration with different negative (\textbf{C}) and positive (\textbf{D})  mixing ratios ($A^0_4/A^0_6$) when compensated with a $Z^0_4$ mode for R/w=1. \textbf{G} Power recovery for a combined $Z^0_6$+$Z^0_8$ aberration when compensated with a two $Z^0_4$ modes of varying R/w. The values on the diagonal are equivalent to using only one $Z^0_4$ mode ($P^1$). The scale is normalised to the maximum of the diagonal ($P^1_{max}$) to show the achievable improvement in power recovery when using two $Z^0_4$ modes.} 
\label{fig:2}
\end{figure*}

\section{Direct evaluation of Zernike aberration cross-coupling in Gaussian beams}

The framework based on analysing the coupling of aberrated Gaussian beams into higher order LG-beams is very useful, however in this particular application it suffers from the limitations due to the uncertainty of evaluating the crossangle (\textbf{Figure \ref{fig:1}D}). To alleviate this problem we propose an alternative formulation based on directly evaluating the crosscoupling between Zernike aberrated Gaussian beams \cite{mafusire2020zernike}:

\begin{equation}
    k_{GG}^{n,m,n',m'}=\int_A G\exp(ikZ^m_n)\exp(ikZ^{m'}_{n'})^*G^*dA =\int_A G^2\exp[ik(Z^m_n-Z^{m'}_{n'})]dA
    \label{eq:kGG_2Z}
\end{equation}

in the weak aberration regime we can approximate:  

\small
\begin{equation}
    \exp[ik(Z^m_n-Z^{m'}_{n'})]\approx1+ik(Z^m_n-Z^{m'}_{n'})+\frac{1}{2}[ik(Z^m_n-Z^{m'}_{n'})]^2
    \label{eq:aprox}
\end{equation}
\normalsize

where we take into consideration the 2nd order due to evaluating self-coupling (Gaussian mode to Gaussian mode) which is stronger than cross-coupling. By putting \textbf{Equation \ref{eq:aprox}} into \textbf{Equation \ref{eq:kGG_2Z}} we get:



\small
\begin{equation}
    k_{GG}^{n,m,n',m'}=\overset{=1}{\int_A G^2dA}+\overset{\mathbf{I_{G2Z}}}{ik\int_A G^2Z^m_ndA} - \overset{\mathbf{I_{G2Z'}}}{ik\int_A G^2Z^{m'}_{n'}dA} -\overset{\mathbf{I_{G2Z2}}}{\frac{k^2}{2}\int_A G^2(Z^m_n)^2dA} -\overset{\mathbf{I_{G2Z'2}}}{\frac{k^2}{2}\int_A G^2(Z^{m'}_{n'})^2dA}
    +\overset{\mathbf{I_{G2ZZ'}}}{k^2\int_A G^2Z^{m}_{n}Z^{m'}_{n'}dA}
\end{equation}
\normalsize

where $I_{G2Z}$ and $I_{G2Z'}$ can be evaluated using \textbf{Equation \ref{eq:knmplpl2}}, $I_{G2Z2}$ and $I_{G2Z'2}$ can be evaluated using \textbf{Equation S14}. This leaves only $I_{G2ZZ'}$ to be derived (see \textbf{Supplement 1 Section S5}). \textbf{Figure \ref{fig:2}A} shows how this framework can be used to evaluate the self-compensation dependence of $Z^0_4$ (spherical aberration) on the R/w ratio using power recovery (see \textbf{Supplement 1 Section S7} for details). This framework can then be extended to accommodate arbitrary combinations of aberrations:

\small
\begin{equation}
    k_{GG}^Z=\int_A G^2 \exp\bigg(ik\sum_\alpha Z_\alpha\bigg)dA \approx 1 + ik\sum_\alpha I_\alpha-\frac{k^2}{2}\sum_{\alpha,\beta} I_{\alpha,\beta}
\end{equation}
\normalsize

where $I_\alpha=\int_A G^2Z_\alpha dA$ and $I_{\alpha,\beta}=\int_A G^2Z_\alpha Z_\beta dA$. As we use the power coupled into the G-mode as a quality metric for beam aberrations we can explicitly calculate the power coupled into the fundamental G-mode:
    
\begin{equation}
    |k_{GG}^Z|^2=1-k^2\sum_{\alpha,\beta} I_\alpha I_\beta-\frac{k^2}{2}\sum_{\alpha,\beta} I_{\alpha,\beta}+\frac{k^4}{4}\sum_{\alpha,\beta,\gamma,\delta}I_{\alpha,\beta}I_{\gamma,\delta}
    \label{eq:k_GG^Z}
\end{equation}

One conclusion following from \textbf{Equation \ref{eq:k_GG^Z}} is that the power coupling dependence on aberration amplitude in the weak aberration regime is described by an n-dimensional 4th order polynomial:

\small
\begin{equation}
    p(\mathbf{A})=\sum_{(i,j,k,l)}\sum_{\alpha=0}^4 \sum_{\beta=0}^{4-\alpha}\sum_{\gamma=0}^{4-\alpha-\beta} \sum_{\delta=0}^{4-\alpha-\beta-\gamma} A_i^\alpha A_j^\beta A_k^\gamma A_l^\delta c_{\alpha,\beta,\gamma,\delta}^{i,j,k,l}
\end{equation}
\normalsize

where (i,j,k,l) are sets of distinct indices. For practical applications we differentiate aberrations into passive aberrations (the aberrations present in the system which we cannot control) and active aberrations (the modes our AO element can display in a controlled manner). Using this approach we investigated two simple cases of 1 and 2 active aberrations in a background of passive aberrations. In these cases the polynomials simplify to:

\small
\begin{equation}
\label{eq:P1}
    P^1(A_i,\mathbf{A_{passive}})=\sum_{\alpha=0}^4 A_i^\alpha c_{\alpha}(\mathbf{A_{passive}})
\end{equation}
\normalsize

and

\small
\begin{equation}
\label{eq:P2}
    P^2(A_i,A_j,\mathbf{A_{passive}})=\sum_{\alpha=0}^4 \sum_{\beta=0}^{4-\alpha} A_i^\alpha A_j^\beta c_{\alpha,\beta}(\mathbf{A_{passive}})
\end{equation}
\normalsize

where $c_\alpha$ and $c_{\alpha,\beta}$ are the polynomial coefficients (see \textbf{Supplement 1 Section S6} for explicit form). We used \textbf{Equation \ref{eq:P1}} to show that a low order aberration ($Z^0_4$) is capable of significantly compensating aberrations even several orders higher (such as $Z^0_{10}$) when an appropriate R/w ratio is used (\textbf{Figure \ref{fig:2}B}). Important to note is the fact that $Z^0_{4}$ shows significant cross-compensation for $Z^0_{6}$ even at R/w=1 but for higher orders the cross-compensation at R/w=1 is negligible which is in line with our previous results using the LG mode framework \cite{Czuchnowski:21}.

We have also explored more complex systems of aberrations by using \textbf{Equation \ref{eq:P2}} to evaluate how self-compensation influences cross-compensation by modeling the interaction between a $Z^0_{4}+Z^0_{6}$ aberration and a $Z^0_{4}$ correction (\textbf{Figure \ref{fig:2}C,D}). We can appreciate that the behaviour of the system changes depending on the interaction between the passive aberrations. For a negative amplitude ratio ($A^0_4/A^0_6$) there is a smooth transition between the states, with an expected reduction in power recovery for aberrations with a higher content of $Z^0_{6}$ (as self-compensation is more efficient than cross-compensation, \textbf{Figure \ref{fig:2}C}). However, for positive amplitude ratios the situation is drastically different (\textbf{Figure \ref{fig:2}D}), there is no gradual transition but more of a bi-stable situation where the system seem to switch between 2 optimal R/w ratios. To understand this behaviour better we need to study how the power recovery evolves for R/w=1. We observe that for negative ratios the power recovery slowly decreases, which implies there is no cross-compensation in the background $Z^0_{4}+Z^0_{6}$ aberration (\textbf{Figure \ref{fig:2}E}). On the other hand for positive ratios the power recovery has a minimum at around $A^0_4/A^0_6=1/4$ (\textbf{Figure \ref{fig:2}E}) which indicates that the background aberrations are cross-compensating and nearly balanced for $A^0_4/A^0_6=1/4$ in a way where modifying the $A^0_4$ via aberration correction does not yield a significant improvement. This leads to a bi-stable behaviour where if the $Z^0_4$ contribution dominates the most efficient strategy is to compensate $Z^0_4$ by using R/w=1, but if $Z^0_6$ is dominant it is more efficient to compensate the overall aberration by using $R/w\approx0.7$.

Finally, by analysing \textbf{Figure \ref{fig:2}B} one can begin to hypothesise that using two $Z^0_4$ aberration at different R/w ratios ($R\approx0.7$ and $R\approx0.5$, which in practice would require two separate DMs) might allow efficient simultaneous compensation of a $Z^0_{6}+Z^0_{8}$ aberration further extending the compensation range of DMs. We have tested this hypothesis by again using \textbf{Equation \ref{eq:P2}} and calculated the expected power recovery for different combinations of R/w ratios used (\textbf{Figure \ref{fig:2}G}). We note that while the increase in power recovery compared to using 1 DM is only moderate ($\sim 20\%$) it is striking that it is possible to extend the power recovery of modes 2 orders higher that the one used for correction ($Z^0_{4}$ vs $Z^0_{8}$).

\section{Discussion}

Over the past years, the concept of adaptive optics has developed into a powerful method to counteract optical aberrations which allowed for a multitude of imaging related applications most notably in astronomy, but also in biology. 
However, hardware limitations of typical AO elements such as deformable mirrors in terms of number of active elements still prohibits robust aberration correction of higher order Zernike aberrations. Our theoretical work explores the possibility of enhancing the capabilities of already existing DMs, therefore allowing efficient correction of higher-order Zernike aberrations using only lower-order modes. We show that it is indeed possible to partially correct aberrations even 3 orders higher than the mode used for correction, as well as to correct several aberrations at the same time. On a practical note, it is important to consider that due to the cross-compensation of background aberration this proposed framework will work best when two AO elements are employed, one operating at R/w=1 to compensate the lower-order aberration within the limits of the AO element and another one working at  R/w<1 to compensate for higher order aberrations. Finally, we believe that this work will enable further studies in the direction of enhancing the capabilities of existing AO devices to tackle increasingly higher order aberrations, which could have significant impact on challenging astronomy and deep-tissue imaging applications.

\section*{Funding}
This work was supported by the European Molecular Biology Laboratory (EMBL), the Chan Zuckerberg Initiative (Deep Tissue Imaging grant no. 2020-225346), as well as the Deutsche Forschungsgemeinschaft (DFG, project no. 425902099).

\section*{Disclosures}

The authors declare that there are no conflicts of interest related to this article.

\bibliographystyle{naturemag}  
\bibliography{bib}

\newpage
\clearpage

\beginsupplement

{\Large\textbf{{Supplementary Information}}}

This document provides supplementary information on the main manuscript "Zernike mode rescaling extends capabilities of adaptive optics for microscopy". It contains full definitions of functions and derivations important integrals used in our theoretical approach.

\section{Supplementary definitions}

We use the following definitions for Zernike aberrations and Laguerre-Gaussian beams. The Zernike aberrations are defined as:

\begin{equation}
Z^m_n(r,\phi) = \begin{cases}
      A^m_n R^m_n(r)\cos(m\phi)  & \text{for $m\geqslant0$}\\
      A^m_n R^{|m|}_n(r)\sin(|m|\phi) & \text{for $m<0$},\\
    \end{cases}   
\end{equation}

where,

\begin{equation}
R^m_n(r)=\sum \limits_{s=0}^{(n-m)/2} \frac{(-1)^s(n-s)!}{s!((n+m)/2-s)!((n-m)/2-s)!}r^{(n-2s)}.
\label{eq:Rmn}
\end{equation}

And $A^m_n$ is the aberration magnitude. The general LG mode with indices $l,p$ is defined as:

\begin{equation}
{LG_{lp}}(r,\phi ,z,k)=C_{lp}^{LG}\left({\frac {r{\sqrt {2}}}{w(z)}}\right)^{\!|l|}L_{p}^{|l|}\!\left({\frac {2r^{2}}{w^{2}(z)}}\right)\exp(-il\phi )G(r,z,k)
\end{equation}

where $C_{lp}^{LG}$ is a normalisation constant, $L_p^{|l|}(x)$ is the Laguerre polynomial and $G(r,z,k)$ describes a general Gaussian beam:

\begin{equation}
G(r,z,k) = E_{0}\frac {w_{0}}{w(z)}\exp \!\left(\!{\frac {-r^{2}}{w(z)^{2}}}\right) \exp \bigg(-i \bigg(kz+{\frac {kr^{2}}{2R(z)}}-\psi (z) \bigg)\bigg)
\end{equation}

$w(z)$ is the local beam radius:

\begin{equation}
w(z)=w_0\sqrt{1+\bigg( \frac{z}{z_{\mathrm{R}}} \bigg)^2},
\end{equation}

$R(z)$ is the local beam curvature:

\begin{equation}
R(z)=z\bigg[1+\bigg(\frac{z_{\mathrm{R}}}{z} \bigg)^2\bigg],
\end{equation}

$\psi(z)$ is the Gouy phase:

\begin{equation}
\label{eq:Gouy}
\psi(z)=(2p+|l|+1)\arctan \bigg(\frac{z}{z_{\mathrm{R}}} \bigg),
\end{equation}

and $z_{\mathrm{R}}$ the Rayleigh range of the beam:

\begin{equation}
z_{\mathrm{R}}=\frac{\pi {w_0}^2 n_0}{\lambda},
\label{eq:Zr}
\end{equation}

where, $n_0$ is the refractive index of the propagation medium and $w_0$ is the beam radius in focus. 

\section{Azimuthal integral}

The explicit form of the azimuthal integral ($I_\phi$, see ref. \cite{Bond:11,Bond:14,Czuchnowski:21} for details): 

\begin{equation}
\label{eq:phi}
[I_\phi]^{n,m}_{p,l,p',l'} =
    \begin{cases}
      0 & \text{if $m \neq |l-l'|$}\\
      \pi & \text{if $m=|l-l'|$, \ even $Z^m_n$}\\
      sgn(l-l') i\pi & \text{if $m=|l-l'|$, \ odd $Z^m_n$}\\
      2\pi & \text{if $m=0$, \ $l=l'$}
    \end{cases}   
\end{equation}

\section{Lower bound power recovery}

The lower bound of power recovery assumes that that the energy contained in high-order LG modes (beyond the mode number considered in the calculation) is not compensated by the correction which lowers the overall recoverable power:

\small
\begin{equation}
   P_{rec}^{lower}=\frac{\sum_i \big(k_i^{n_p,m_p}(\frac{R_p}{w})+\alpha k_i^{n_a,m_a}(\frac{R_a}{w})\big)\overline{\big(k_i^{n_p,m_p}(\frac{R_p}{w})+\alpha k_i^{n_a,m_a}(\frac{R_a}{w})\big)}+P_{missing}^{passive}+P_{missing}^{active}}{1-\big|k_{0,0,0,0}^{n_p,m_p}(\frac{R_p}{w})\big|^2}
\end{equation}
\normalsize

where:

\small
\begin{equation}
   \alpha=-\frac{\sum_i \big(k_i^{n_a,m_a}(\frac{R_a}{w}) \overline{k_i^{n_p,m_p}(\frac{R_p}{w}})+k_i^{n_p,m_p}(\frac{R_p}{w}) \overline{k_i^{n_a,m_a}(\frac{R_a}{w}})\big)}{2\sum_i |k_i^{n_a,m_a}(\frac{R_a}{w})|^2},
\end{equation}
\normalsize

\begin{equation}
   P_{missing}^{passive}=1-\big|k_{0,0,0,0}^{n_p,m_p}(\frac{R_p}{w})\big|^2-\sum_i \big|k_i^{n_p,m_p}(\frac{R_p}{w})\big|^2
\end{equation}

and

\begin{equation}
   P_{missing}^{active}=1-\big|k_{0,0,0,0}^{n_a,m_a}(\frac{R_a}{w})\big|^2-\alpha^2\sum_i \big|k_i^{n_a,m_a}(\frac{R_a}{w})\big|^2.
\end{equation}

where $i=(p,l,p',l')$ for $(p',l')\neq (0,0)$.

\section{Expression for the I\textsubscript{G2Z2} integral}

The expression for the $I_{G2Z2}$ integral can be directly calculated from the expression derived in ref. \cite{Bond:14} and expressed explicitly in ref. \cite{Czuchnowski:21}:

\begin{multline}
   I_{G2Z2} =(A^m_n)^2\frac{I_{\phi}^{m}}{2\pi} \sum_{h=0}^{\frac{1}{2}(n-m)} \sum_{g=0}^{\frac{1}{2}(n-m)} \frac{(-1)^{h+g}(n-h)!X^{h+g-n}}{(\frac{1}{2}(n+m)-h)!(\frac{1}{2}(n-m)-h)!h!}\times \\
    \times \frac{(n-g)!}{(\frac{1}{2}(n+m)-g)!(\frac{1}{2}(n-m)-g)!g!}\gamma(n-h-g+1,X)
\label{eq:I_G2Z2}
\end{multline}
 
where:

\begin{equation}
I_\phi^{m}=
    \begin{cases}
      2\pi & \text{if $m = 0$}\\
      1\pi & \text{if $m \neq 0$}\\
    \end{cases}  
\end{equation}

and $X=2R^2/\omega^2$ describes the scaling between the Zernike radius ($R$) size and the beam radius ($\omega$), $A^m_n$ denotes the Zernike aberration amplitude and $\gamma(a,x)=\int^x_0 t^{a-1}e^{-t}dt$ is the lower incomplete gamma function.

\section{Derivation of the analytical expression for the I\textsubscript{G2ZZ'} integral}

\begin{equation}
    I_{G2ZZ'}=A^m_nA^{m'}_{n'}I_{\phi}^{m,m'}\int_0^R \frac{2}{\pi w^2} 
    \exp\bigg(-\frac{2r^2}{w^2}\bigg)R^{|m|}_n\bigg(\frac{r}{R}\bigg)R^{|m'|}_{n'}\bigg(\frac{r}{R'}\bigg)rdr
\end{equation}

assuming $R < R'$, we change the variable by substituting $x=\frac{2r^2}{w^2}$, the integration limit becomes $X=\frac{2R^2}{w^2}$:

\begin{equation}
    I_{G2ZZ'}=A^m_nA^{m'}_{n'}\frac{I_{\phi}^{m,m'}}{2\pi}\int_0^X \exp(-x)R^{|m|}_n\bigg(\frac{\sqrt{x}}{\sqrt{X}}\bigg)R^{|m'|}_{n'}
    \bigg(\frac{R\sqrt{x}}{R'\sqrt{X}}\bigg)dx
\end{equation}

using \textbf{Equation \ref{eq:Rmn}} we can explicitly write:
\begin{multline}
    I_{G2ZZ'}=A^m_nA^{m'}_{n'}\frac{I_{\phi}^{m,m'}}{2\pi}\sum_{h=0}^{\frac{n-m}{2}}\sum_{h'=0}^{\frac{n'-m'}{2}}\frac{(-1)^{h}(n-h)!}{h!(\frac{n+m}{2}-h)!(\frac{n-m}{2}-h)!}\\
    \times \frac{(-1)^{h'}(n'-h')}{h!(\frac{n'+m'}{2}-h')!(\frac{n'-m'}{2}-h')!}\frac{R^{(n'-2h')}}{R'^{(n'-2h')}X^{\frac{1}{2}(n+n'-2h-2h')}} \\ 
    \times \int_0^X \exp(-x)x^{\frac{1}{2}(n+n'-2h-2h')}dx
\end{multline}

which can be simplified using the lower incomplete gamma function $\gamma(a,x)=\int_0^x t^{a-1}e^{-t}dt$:

\begin{multline}
    I_{G2ZZ'}=A^m_nA^{m'}_{n'}\frac{I_{\phi}^{m,m'}}{2\pi}\sum_{h=0}^{\frac{n-m}{2}}\sum_{h'=0}^{\frac{n'-m'}{2}}\frac{(-1)^{h}(n-h)!}{h!(\frac{n+m}{2}-h)!(\frac{n-m}{2}-h)!}\\
    \times \frac{(-1)^{h'}(n'-h')}{h!(\frac{n'+m'}{2}-h')!(\frac{n'-m'}{2}-h')!}\frac{R^{(n'-2h')}}{R'^{(n'-2h')}X^{\frac{1}{2}(n+n'-2h-2h')}} \\ 
    \times \gamma(\frac{1}{2}(n+n')-h-h',X)
\end{multline}

where:

\begin{equation}
I_\phi^{m,m'}=
    \begin{cases}
      \int_0^{2\pi}\cos(m\phi)\cos(m'\phi)d\phi=\pi\epsilon_{m,m'}\delta_{m,m'}, & m,m'\geq0\\
      \int_0^{2\pi}\sin(|m|\phi)\sin(|m'|\phi)d\phi=\pi\delta_{m,m'}, & m,m'<0\\
      \int_0^{2\pi}\sin(|m|\phi)\cos(m'\phi)d\phi=0, & m<0<m'\\
      \int_0^{2\pi}\cos(m\phi)\sin(|m'|\phi)d\phi=0, & m'<0<m\\
    \end{cases}  
\end{equation}

where:

\begin{equation}
\delta_{m,m'}=
    \begin{cases}
      0 & \text{if $m \neq m'$}\\
      1 & \text{if $m = m'$}\\
    \end{cases}  
\end{equation}

and:

\begin{equation}
\epsilon_{m,m'}=
    \begin{cases}
      2 & \text{if $m, m' = 0$}\\
      1 & \text{if $m, m' \neq 0$}\\
    \end{cases}  
\end{equation}

The solution for $R > R'$ is analogous:

\begin{multline}
    I_{G2ZZ'}=A^m_nA^{m'}_{n'}\frac{I_{\phi}^{m,m'}}{2\pi}\sum_{h=0}^{\frac{n-m}{2}}\sum_{h'=0}^{\frac{n'-m'}{2}}\frac{(-1)^{h}(n-h)!}{h!(\frac{n+m}{2}-h)!(\frac{n-m}{2}-h)!}\\
    \times \frac{(-1)^{h'}(n'-h')}{h!(\frac{n'+m'}{2}-h')!(\frac{n'-m'}{2}-h')!}\frac{R'^{(n-2h)}}{R^{(n-2h)}X^{\frac{1}{2}(n+n'-2h-2h')}} \\ 
    \times \gamma(\frac{1}{2}(n+n')-h-h',X)
\end{multline}

\section{Explicit form of coupling polynomials}

The case of only a single active aberration simplifies \textbf{Equation 15} to a 1d 4th-order polynomial:

\begin{equation}
    p(A_i,\mathbf{A_{passive}})=\sum_{\alpha=0}^4 A_i^\alpha c_{\alpha}(\mathbf{A_{passive}})
\end{equation}

where the coefficients $c_\alpha$ can be explicitly written as:

\begin{equation}
    c_{0}=1-k^2\sum_{\alpha,\beta\neq i} I_\alpha I_\beta-\frac{k^2}{2}\sum_{\alpha,\beta\neq i} I_{\alpha,\beta}+\frac{k^4}{4}\sum_{\alpha,\beta,\gamma,\delta\neq i}I_{\alpha,\beta}I_{\gamma,\delta}
\end{equation}

\begin{equation}
    c_{1}=-2k^2\sum_{\alpha\neq i} I_\alpha I_i-k^2\sum_{\alpha\neq i} I_{\alpha,i}+k^4\sum_{\alpha,\beta,\gamma\neq i}I_{\alpha,\beta}I_{\gamma,i}
\end{equation}

\begin{equation}
    c_{2}=k^2I_i^2-\frac{k^2}{2}I_{i,i}+\frac{k^4}{2}\sum_{\alpha,\beta\neq i}[I_{\alpha,\beta}I_{i,i}+2I_{\alpha,i}I_{\beta,i}]
\end{equation}

\begin{equation}
    c_{3}=k^4\sum_{\alpha\neq i}I_{\alpha,i}I_{i,i}
\end{equation}

\begin{equation}
    c_{4}=\frac{k^4}{4}I_{i,i}^2
\end{equation}

The case of two active aberrations simplifies \textbf{Equation 15} to a 2d 4th-order polynomial (important to note is that this form remains the same regardless if the two active aberrations have the same or different R/w ratios):

\begin{equation}
    p(A_i,A_j,\mathbf{A_{passive}})=\sum_{\alpha=0}^4 \sum_{\beta=0}^{4-\alpha} A_i^\alpha A_j^\beta c_{\alpha,\beta}(\mathbf{A_{passive}})
\end{equation}

where the coefficients $c_{\alpha,\beta}$ can be explicitly written as:

\begin{equation}
    c_{00}=1-k^2\sum_{\alpha,\beta\neq i,j} I_\alpha I_\beta-\frac{k^2}{2}\sum_{\alpha,\beta\neq i,j} I_{\alpha,\beta}+\frac{k^4}{4}\sum_{\alpha,\beta,\gamma,\delta\neq i,j}I_{\alpha,\beta}I_{\gamma,\delta}
\end{equation}

\begin{equation}
    c_{10}=-2k^2\sum_{\alpha\neq i,j} I_\alpha I_i-k^2\sum_{\alpha\neq i,j} I_{\alpha,i}+k^4\sum_{\alpha,\beta,\gamma\neq i,j}I_{\alpha,\beta}I_{\gamma,i}
\end{equation}

\begin{equation}
    c_{20}=k^2I_i^2-\frac{k^2}{2}I_{i,i}+\frac{k^4}{2}\sum_{\alpha,\beta\neq i,j}[I_{\alpha,\beta}I_{i,i}+2I_{\alpha,i}I_{\beta,i}]
\end{equation}

\begin{equation}
    c_{30}=k^4\sum_{\alpha\neq i,j}I_{\alpha,i}I_{i,i}
\end{equation}

\begin{equation}
    c_{40}=\frac{k^4}{4}I_{i,i}^2
\end{equation}

\begin{equation}
    c_{11}=2k^2I_i I_j-k^2I_{i,j}+k^4\sum_{\alpha,\beta\neq i,j}[I_{\alpha,\beta}I_{i,j}+2I_{\alpha,i}I_{\beta,j}]
\end{equation}

\begin{equation}
    c_{21}=k^4\sum_{\alpha,\beta\neq i,j}[I_{\alpha,j}I_{i,i}+2I_{\alpha,i}I_{i,j}]
\end{equation}

\begin{equation}
    c_{31}=k^4I_{i,i}I_{i,j}
\end{equation}

\begin{equation}
    c_{22}=\frac{k^4}{2}[I_{i,i}I_{j,j}+2I_{i,j}^2]
\end{equation}

\section{Power recovery}

The power recovery ($P_{rec}$) for both self- and cross-compensation can be defined as follows:

\begin{equation}
    P_{rec}^X=\bigg[1-\frac{1-min\big[|k_{GG}^Z(X_{active},\mathbf{A_{active}})|^2\big]}{1-|k_{GG}^Z(\mathbf{A_{active}=0})|^2}\bigg]\times 100\%
\end{equation}

where $X=\frac{2R^2}{w^2}$ and $\mathbf{A_{active}}$ denotes the set of Zernike amplitudes for the active aberrations (Zernike mode displayed on the AO element and actively controllable).

\newpage
\section{Supplementary Figures}

\begin{figure}[h!]
\includegraphics[width=16cm]{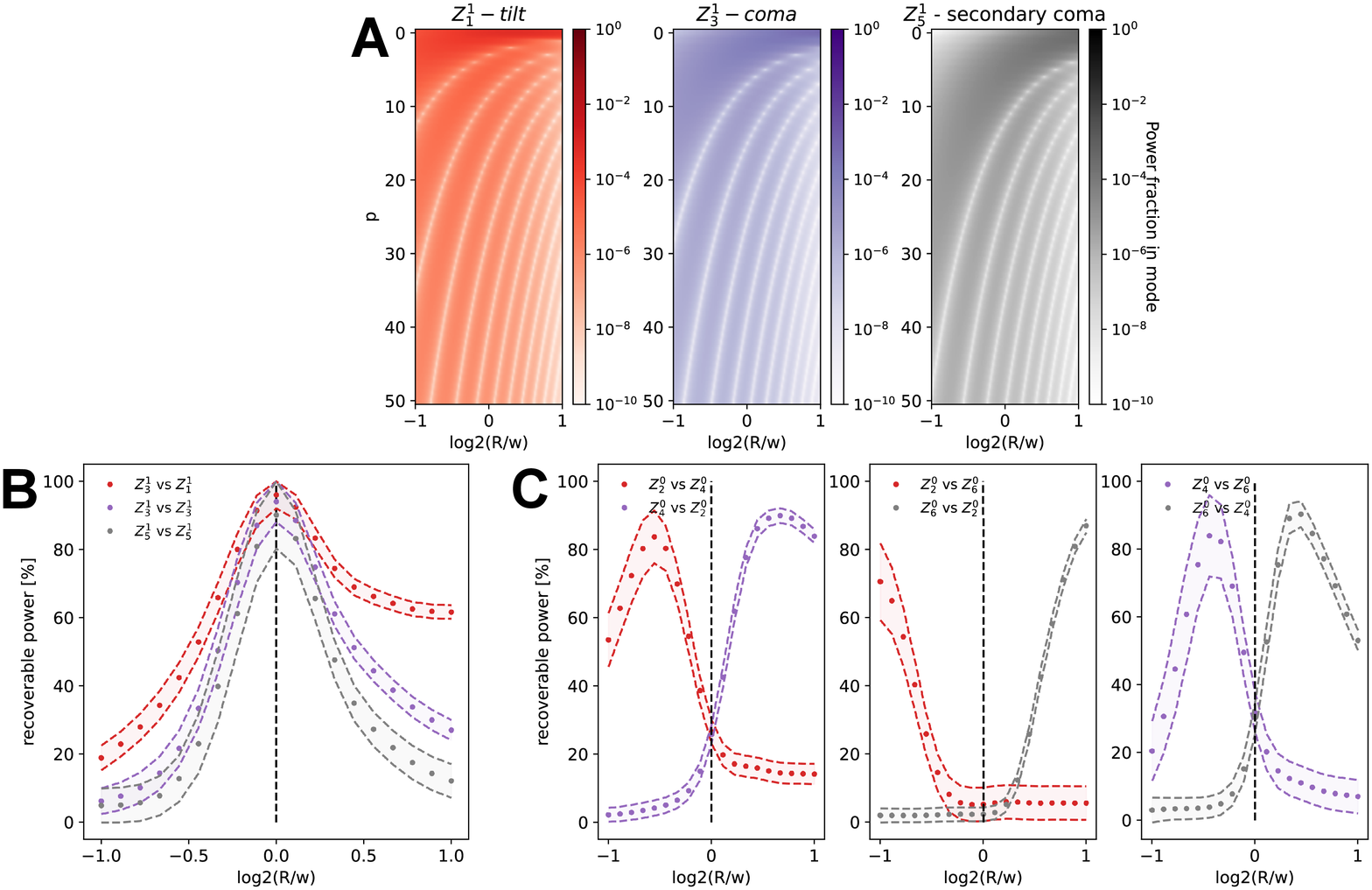}
\centering
\caption{\textbf{A} Dependence of the power coupling into higher order LG-modes on the ratio between the aberration and beam radii (R/w). \textbf{B} Dependence of the self-compensation in LG-space on the R/w ratio using the first 25 LG modes. The shaded area represents the range between the lower and upper bounds calculated using \textbf{Equation 8} and \textbf{Equation S10} (see \textbf{ Section S3} for details). The dots represent the expected value. \textbf{C} Dependence of cross-compensation between different aberrations on the R/w ratio using the first 25 LG modes.} 
\label{fig:Correction of a single aberration with two cross-compensating Zernike modes.}
\end{figure}

\end{document}